\documentclass[dvipdfmx, letterpaper, 10 pt, conference]{ieeeconf}
\IEEEoverridecommandlockouts
\usepackage{amsmath,amssymb}
\newtheorem{theorem}{Theorem}
\newtheorem{proposition}{Proposition}
\newtheorem{lemma}{Lemma}
\newtheorem{definition}{Definition}

\newtheorem{corollary}{Corollary}
\usepackage{algorithm, algpseudocode}
\algrenewcommand\algorithmicrequire{\textbf{Input:}}
\algrenewcommand\algorithmicensure{\textbf{Output:}}
\usepackage{graphicx}
\usepackage{cite}
\DeclareMathOperator{\tr}{tr}
\DeclareMathOperator{\diag}{diag}
\DeclareMathOperator{\E}{E}
\DeclareMathOperator{\Cov}{Cov}
\DeclareMathOperator*{\minimize}{minimize}
\DeclareMathOperator*{\maximize}{maximize}
\DeclareMathOperator*{\argmax}{arg\,max}
\newcommand{\e}{\text{e}}
\newcommand{\gammamin}{\underline{\gamma}}
\newcommand{\alphamax}{\overline{\alpha}}
\newcommand{\lambdamin}{\lambda_{\min}}
\newcommand{\lambdamax}{\lambda_{\max}}

\newif\ifshowonlyref
\usepackage{mathtools}
\mathtoolsset{showonlyrefs=true}
\showonlyreftrue

\title{\LARGE \bf
Sensor placement minimizing the state estimation mean square error:
 Performance guarantees of greedy solutions}
 
\author{Akira Kohara, Kunihisa Okano, Kentaro Hirata, and Yukinori Nakamura%
 \thanks{The authors are with the Department of Intelligent Mechanical Systems,
 Graduate School of Natural Science and Technology, Okayama University,
 Okayama, 700-8530, Japan. E-mails: 
 {\footnotesize \{\texttt{akirak@s.}, \texttt{kokano@}, \texttt{kent@sys.},
 \texttt{yukinori-n@}\}\texttt{okayama-u.ac.jp}}.}%
 \thanks{This work was supported by JSPS KAKENHI Grant Numbers
 18K13778 and 20K14763.}%
}

\begin{document}
\maketitle
\thispagestyle{empty}
\pagestyle{empty}

\begin{abstract}
This paper studies selecting a subset of the system's output to minimize
the state estimation mean square error (MSE).
This results in the maximization problem of a set function defined on possible sensor
selections subject to a cardinality constraint.
We consider to solve it approximately by a greedy search.
Since the MSE function is not submodular nor supermodular, the well-known performance
guarantees for the greedy solutions do not hold in the present case.
Thus, we use the quantities---the submodularity ratio and the curvature---to
evaluate the degrees of submodularity and supermodularity of the objective function.
By using the properties of the MSE function, we approximately compute these quantities
and derive a performance guarantee for the greedy solutions.
It is shown that the guarantee is less conservative than those in the existing results.
\end{abstract}

\section{Introduction}\label{sec:intro}
In the last few years, a considerable number of studies have been made on
control and state estimation of large-scale complex systems such as
power systems \cite{Pasqualetti2014} and biochemical networks \cite{Liu2011b, Liu2013}.
For such systems, installing sensors and actuators onto all possible ports
would be impossible due to a budget constraint and limitations on power and communication
resources.
Thus, there is a need to establish a method to determine which information should be
observed/actuated in the course of designing systems.

This paper concerns the sensor placement problem.
Our objective is to minimize the state estimation error with a given (integer) number
of the sensors.
It is known that this problem is modeled as the maximization of a set function and
it is in general NP-hard (see, e.g., \cite{Zhang2017,Olshevsky2014,Tzoumas2016b,Ye2018}).
Hence, many studies have attempted to solve the problem approximately.
One of the major approaches is to employ a continuous relaxation and
reduce the problem to a convex optimization as in \cite{Joshi2009}.

Another approach is to use greedy algorithms
\cite{Shamaiah2010,Summers2016,Zhang2017,Tzoumas2016,Chamon2017,Chamon2020a}:
Pick the sensor which achieves
the largest increment of the objective function one by one until the number of 
selected sensors reaches the upper bound.
In addition to its simplicity in implementation, a celebrated feature of the algorithms
is the performance guarantees.
When the objective function is submodular \cite{Shamaiah2010,Summers2016,Tzoumas2016},
the ratio between the optimum and the value of the function for the greedy solution is
theoretically guaranteed \cite{Nemuhauser1978}.

The class of \emph{non-}submodular functions, however, contains an important function;
the state estimation mean square error (MSE).
Several papers have addressed performance guarantees in the minimization of MSE
\cite{Summers2019a,Chamon2017,Chamon2020a}
\footnote{We note that the result corresponding to the MSE in \cite{Summers2016} has been
corrected in \cite{Summers2018}.}.
The authors have employed quantities to evaluate how far the MSE is from being submodular:
Summers and Kamgarpour \cite{Summers2019a} have used the submodularity ratio
\cite{Das2011} and the curvature \cite{Conforti1984,Bian2017},
while Chamon \textit{et al.}\ \cite{Chamon2017,Chamon2020a} have applied the notion of
approximate submodularity \cite{Chamon2016}, which is called $a$-modularity in
\cite{Lehmann2006}.

In this paper, we also employ the state estimation MSE as the objective function.
We address the smoothing problem: Given a bunch of the outputs for a time
period, estimate the states during the period.
The MSE for this problem is evaluated based on the submodularity ratio and the
curvature as in \cite{Summers2019a}.
We show that by employing a less conservative evaluation of the quantities,
we obtain a tighter guarantee for the greedy algorithm than those in the existing work.

The paper is organized as follows.
In Section~\ref{sec:problem}, we formally describe the considered problem.
Next, we give preliminary results on the maximization of set functions in Section
\ref{sec:setfunction}.
We then present the main result of the paper in Section \ref{sec:main} and compare
it with the existing work in Section \ref{sec:comparison}.
Finally, concluding remarks are provided in Section \ref{sec:conclusion}.

\paragraph*{Notation}
Throughout the paper, $\mathbb{R}$ denotes the set of real numbers, and
$\mathbb{Z}_+$ is the set of nonnegative integers.
Furthermore, %
$0$ denotes the zero vector or matrix of appropriate size.
The matrix $\diag(d_1,d_2,\dots,d_n)$ is the diagonal matrix where
the diagonal elements are $d_i$s.
We use the same notation for the block diagonal matrix with matrices
$d_i$s.
For a matrix $A$, $[A]_{i,j}$ denotes the $(i,j)$ element of $A$.
The symbol $\otimes$ is the Kronecker product.
For a finite set $\mathcal{X}$, $|\mathcal{X}|$ and $2^{\mathcal{X}}$ represent
the cardinality and the power set of $\mathcal{X}$, respectively.
Finally, for a random variable (vector) $\xi$, $\E[\xi]$ and $\Cov[\xi]$ denote
the expectation and the variance (covariance matrix) of $\xi$, respectively.

\section{Sensor placement problem}\label{sec:problem}
In this section, we first introduce a linear dynamical system and explain
the considered state estimation problem.
We then formulate the optimal sensor placement problem for the estimation.
Let us consider the following linear time-invariant system:
\begin{subequations}\label{system1}
\ifshowonlyref
 \noeqref{system1a,system1b}
\fi
\begin{align}
 x_{k+1} &= Ax_k + w_k,\label{system1a}\\
 y_k &= C x_k + v_k,\label{system1b}
\end{align}
\end{subequations}
where $x_k \in \mathbb{R}^n$ and $y_k \in \mathbb{R}^p$ are the state and the
output of the system at time $k \in \mathbb{Z}_+$, respectively.
The initial state $x_0$ follows a Gaussian distribution with mean $\E[x_0]$ and
covariance $\Cov[x_0]=X_0 \succ 0$.
The disturbance $w_k \in \mathbb{R}^n$ and the noise $v_k \in \mathbb{R}^p$ are
zero-mean Gaussian vectors with covariances $\Cov[w_k]=W\succ 0$ and $\Cov[v_k]
= V := \diag(\sigma_{v,1}^2,\sigma_{v,2}^2, \dots,\sigma_{v,p}^2)\succ 0$ for all
$k\in\mathbb{Z}_+$.
We assume that $w_k,w_{k'},v_k,v_{k'}$ are mutually independent for any
$k,k'\in\mathbb{Z}_+,\ k\neq k'$.

We consider the scenario that at most $s$ ($s\leq p$) elements of the output
$y_k\in\mathbb{R}^p$ are available for the state estimation.
Let the index set of the output be $\mathcal{I}:=\{1,2,\dots,p\}$ and
the set of selected indices $\theta_i$ be
\begin{align}
 &\mathcal{S}
 := \{\theta_1, \theta_2, \dots, \theta_s : \theta_i \in \mathcal{I},\\
 &\quad\quad\quad \theta_i < \theta_{i+1}, i=1,2,\dots,s-1 \}.
\end{align}
Then, the selected output can be written as
$y_{\mathcal{S},k}:=S_\mathcal{S}y_k$
with the selection matrix $S_\mathcal{S}\in\mathbb{R}^{s\times p}$
defined as
\begin{align}
 &\left[S_\mathcal{S}\right]_{i,j} =
 \begin{cases}
  1 & \text{if } j=\theta_i,\\
  0 & \text{otherwise}.
 \end{cases}
\end{align}
We note that the sensor selection $\mathcal{S}$ is time invariant.
that is, once a subset of $y_k$ is selected and the sensors are installed,
they are fixed during operation.

Our objective is to find a sensor selection minimizing the estimation MSE
for the time period $[0,\ell-1]\subset \mathbb{Z}_+$.
We examine the smoothing problem, that is, the estimate $\tilde{x}_k$ of
$x_k$ ($k=0,1,\dots, \ell-1$) is computed from the selected outputs
$\left\{y_{\mathcal{S},k}\right\}_{k=0}^{\ell-1}$ for the whole period.
The objective function, the MSE, is given as
\begin{align}
 J(\mathcal{S}) = \min_{\left\{\tilde{x}_k\right\}_{k=0}^{\ell-1}}
 \sum_{k=0}^{\ell-1}
 \E\left[\left\|x_k-\tilde{x}_k\right\|_2^2
 \big| \left\{y_{\mathcal{S},j}\right\}_{j=0}^{\ell-1}\right].\label{mmse}
\end{align}

We now formally state the sensor placement problem as
\begin{subequations}\label{prob:orig}
\ifshowonlyref
 \noeqref{origa,origb}
\fi
 \begin{alignat}{2}
  &\minimize_{\mathcal{S}\subseteq\mathcal{I}} &\quad &J(\mathcal{S}),\label{origa}\\
  &\text{subject to} &\quad &|\mathcal{S}| \leq s.\label{origb}
 \end{alignat}
\end{subequations}
Problem \eqref{prob:orig} is an NP-hard combinatorial optimization problem,
and thus solving this problem by a brute force search is computationally
demanding.

\subsection{Explicit form of the objective function}
Let us look at the objective function \eqref{mmse}.
We here present an explicit form of the mean square error \eqref{mmse}
by following the arguments in \cite{Tzoumas2016, Chamon2017}.
Combining the output equations 
$y_{\mathcal{S},k} = S_{\mathcal{S}}(C x_k + v_k)$ for $k=0,1,\dots,\ell-1$,
we have
\begin{align}
 \bar{y} = G\bar{z} + \bar{v},\label{lineq}
\end{align}
where 
$\bar{y}:=[y_{\mathcal{S},0}^\top\;\; y_{\mathcal{S},1}^\top\;\;
 \dots\;\; y_{\mathcal{S},\ell-1}^\top]^\top$,
$\bar{z}:=[x_0^\top\;\; w_0^\top\;\; \dots \;\; w_{\ell-2}^\top]^\top$,
$\bar{v}:=[v_{\mathcal{S},0}^\top\;\; v_{\mathcal{S},1}^\top\;\;
\dots \;\; v_{\mathcal{S},\ell-1}^\top]^\top$, 
$v_{\mathcal{S},k}:=S_\mathcal{S}v_k$,
and
\begin{align}
 G:=\left\{I_{\ell}\otimes (S_\mathcal{S}C)\right\}\Phi,\;
 \Phi:=
 \begin{bmatrix}
  I_n & 0 & \cdots &  0\\
  A   & I_n & \ddots &  0\\
  \vdots & \vdots &  \ddots & \vdots\\
  A^{\ell-1} & A^{\ell-2} & \cdots &  I_n
 \end{bmatrix}.
\end{align}
A solution of the minimization in the right-hand side of \eqref{mmse} can be
obtained from a least mean square estimate $\tilde{z}$ of $\bar{z}$ and
\eqref{system1} \cite{Kailath2000}.
Since $\bar{z}$ and $\bar{v}$ are Gaussian, we have the explicit form
of $\tilde{z}$ as 
\begin{align}
 \tilde{z} = ZG^\top \left(V_{\mathcal{S}} + GZG^\top \right)^{-1}\bar{y},
\end{align}
where $Z:=\Cov[\bar{z}]=\diag(X_0, W,\dots, W)$ is the covariance of $\bar{z}$,
and $V_{\mathcal{S}}:=\Cov[\bar{v}]
=I_{\ell}\otimes (S_{\mathcal{S}}VS_{\mathcal{S}}^\top)$
is the covariance of $\bar{v}$.
Furthermore, the covariance of the minimum estimation error is given by
\begin{align}
 \E\left[(\bar{z}-\tilde{z})(\bar{z}-\tilde{z})^\top\right]
 = \left(Z^{-1} + G^\top V_{\mathcal{S}}^{-1} G\right)^{-1}.
 \label{Sigma}
\end{align}
Note that we have $Z\succ 0$ and $V_{\mathcal{S}}\succ 0$ from the
setup.

We see in \eqref{Sigma} that the impact of sensor placement in the estimation
error is represented by the term $G^\top V_{\mathcal{S}}^{-1} G$.
For simplicity of notation, let $U_{\mathcal{S}}:=G^\top V_{\mathcal{S}}^{-1} G$,
and let $L:=Z^{-1}$.
Then, the objective function defined in \eqref{mmse} can be expressed as
\begin{align}
 J(\mathcal{S}) = \tr\left[\left(L+U_{\mathcal{S}}\right)^{-1}\right].
 \label{trSigma}
\end{align}

The following lemma gives an important property of $U_{\mathcal{S}}$.
\begin{lemma}[\! \cite{Tzoumas2016}]\label{lem:Us}
For any selection set $\mathcal{S} \subseteq \mathcal{I}$,
\begin{align}
 U_{\mathcal{S}}
 = \sum_{i\in\mathcal{S}} \sigma_{v,i}^{-2} \Phi^\top
 \left(I_{\ell}\otimes C\right)^\top \left(I_{\ell}\otimes I^{(i)}\right)
 \left(I_{\ell}\otimes C\right)\Phi \succeq 0.
 \label{Usdecompose}
\end{align}
Here, $I^{(i)}$ is the $p\times p$ matrix where $[I^{(i)}]_{i,i}=1$
and the other elements are zero.
\end{lemma}

This lemma implies that $U_{\mathcal{S}}$ can be decomposed to the sum of
the symmetric matrices corresponding to selected sensors.
We emphasize that each summand is positive semidefinite.
This property will be used in the derivation of the main result.

\section{Greedy algorithm for the maximization of (non-)submodular functions}
\label{sec:setfunction}
In this section, we provide preliminary results on optimization of set functions.
In below, we consider the following problem instead of Problem \eqref{prob:orig}:
\begin{subequations}\label{prob:maxf}
\ifshowonlyref
 \noeqref{maxfa,maxfb}
\fi
 \begin{alignat}{2}
 &\maximize_{\mathcal{S}\subseteq\mathcal{V}}&\quad &f(\mathcal{S}),\label{maxfa}\\
 &\text{subject to} &\quad &|\mathcal{S}|\leq s,\label{maxfb}
 \end{alignat}
\end{subequations}
where $\mathcal{V}$ is a discrete set, $f:2^{\mathcal{V}}\to \mathbb{R}$ is a
set function, and $s$ is a given integer.
Notice that we examine maximization in \eqref{maxfa} while minimization is considered
in \eqref{origa}.

One of the most common approaches among the approximate methods for the above
problem is greedy search.
A greedy algorithm for the above problem is given in Algorithm~\ref{alg:greedy}.
Besides its simplicity and good performance, a particular advantage of this method
is that for a class of set functions, we have a theoretical bound on
the deviation of greedy solutions from the optimal.
\begin{algorithm}[tb]
 \caption{A greedy algorithm for Problem \protect\eqref{prob:maxf}}
 \label{alg:greedy}
  \begin{algorithmic}%
   \Require $\mathcal{V}, f, s$ 
   \State $\mathcal{S}_0 \leftarrow \emptyset$
   \For{$i=1,2,\dots, s$}
    \State $\displaystyle
      \omega^* \leftarrow
      \argmax_{\omega\in\mathcal{V}\backslash\mathcal{S}_{i-1}}
      \left[f(\mathcal{S}_{i-1} \cup \left\{\omega\right\})
      - f(\mathcal{S}_{i-1})\right]$
   \smallskip
    \State $\displaystyle
      \mathcal{S}_i \leftarrow \mathcal{S}_{i-1} \cup \left\{\omega^*\right\}$
   \EndFor
   \Ensure $\mathcal{S}^{\text{g}} \leftarrow \mathcal{S}_s$
  \end{algorithmic}
\end{algorithm}

\subsection{Submodular case}
We give the following definitions on set functions.
\begin{definition}%
\label{def:nondecreasing}
 A set function $f:2^{\mathcal{V}}\to \mathbb{R}$ is called \emph{nondecreasing}
if for all subsets $\mathcal{S}_1,\mathcal{S}_2$ satisfying
$\mathcal{S}_1\subseteq\mathcal{S}_2\subseteq\mathcal{V}$ it follows that
\begin{align}
 f(\mathcal{S}_1) \leq f(\mathcal{S}_2).
\end{align}
\end{definition}
\medskip

\begin{definition}%
\label{def:submodular}
 A set function $f:2^{\mathcal{V}}\to \mathbb{R}$ is called \emph{submodular}
if for all subsets $\mathcal{S}_1,\mathcal{S}_2$
($\mathcal{S}_1\subseteq\mathcal{S}_2\subseteq\mathcal{V}$) and
for all $\omega\in\mathcal{V}\backslash\mathcal{S}_2$ it holds that
\begin{align}
 f(\mathcal{S}_1\cup\left\{\omega\right\}) - f(\mathcal{S}_1) \geq
 f(\mathcal{S}_2\cup\left\{\omega\right\}) - f(\mathcal{S}_2).
 \label{submodular}
\end{align}
We say $f$ is \emph{supermodular} if the reversed inequality in \eqref{submodular}
holds, and $f$ is \emph{modular} if \eqref{submodular} holds with equality.
\end{definition}

For a nondecreasing and submodular set function, a guarantee for the approximate
performance of the greedy algorithm has been known.
Let $\mathcal{S}^\text{g}$ denote the solution of Problem \eqref{prob:maxf} obtained
by Algorithm~\ref{alg:greedy}, and let $\mathcal{S}^*$ be the optimal solution.
We refer to the following classical result.
\begin{proposition}[\! \cite{Nemuhauser1978}]\label{prop:Nemhauser}
Let $f$ in Problem \eqref{prob:maxf} be nondecreasing and submodular.
Then, it holds that
\begin{align}
 f(\mathcal{S}^{\text{g}})-f(\emptyset) \geq
 \left(1-\e^{-1}\right)(f(\mathcal{S}^*)-f(\emptyset)).
\end{align}
\end{proposition}
\medskip

Suppose that the objective function is normalized as $f(\emptyset) = 0$.
For such a case, from the proposition we have that $f(\mathcal{S}^{\text{g}})$
is at least $(1-\e^{-1}) \approx 0.63$ times the optimum value $f(\mathcal{S}^*)$.

\subsection{Non-submodular case}
Unfortunately, 
the mean square error $J$ in \eqref{mmse} or \eqref{trSigma} is not
submodular nor supermodular as discussed in \cite{Chamon2017,Summers2018,Summers2019a}.
Hence, Proposition \ref{prop:Nemhauser} is not applicable in our case.

In a recent study \cite{Bian2017}, a theoretical performance bound for non-submodular
cases is provided with an extended notion of submodularity.
To introduce this result, let us define the increment of $f$ by adding
$\Omega\subseteq\mathcal{V}$ to $\mathcal{S}\subseteq\mathcal{V}$ as
\begin{align}
 \rho_{\Omega}(\mathcal{S}) := f(\mathcal{S}\cup\Omega) - f(\mathcal{S}).
\end{align}

The following notions are the keys to characterize non-submodular functions.
\begin{definition}[\! \cite{Das2011}]
 The \emph{submodularity ratio} of a nonnegative set function $f$ is the largest scalar
$\gamma$ such that
\begin{align}
 \sum_{\omega \in \Omega\backslash\mathcal{S}} \rho_{\left\{\omega\right\}}(\mathcal{S})
 \geq \gamma \rho_{\Omega}(\mathcal{S}),\quad
 \forall \Omega, \mathcal{S}\subseteq \mathcal{V}.
 \label{gamma_def}
\end{align}
\end{definition}

\begin{definition}[\!\cite{Bian2017}]
 The \emph{curvature} of a nonnegative set function $f$ is the smallest scalar
$\alpha$ such that
\begin{align}
 &\rho_{\left\{j\right\}}(\mathcal{S}\,\backslash\left\{j\right\}\cup\Omega)
 \geq (1-\alpha)\rho_{\left\{j\right\}}(\mathcal{S}\,\backslash\left\{j\right\}),\\
 &\forall \Omega, \mathcal{S} \subseteq \mathcal{V},
 \quad \forall j\in\mathcal{S}\,\backslash\Omega.
 \label{alpha_def}
\end{align}
\end{definition}
\medskip

It is worth citing Remarks~1, 2 in \cite{Bian2017} to provide an intuitive
explanation for these notions:
For a nondecreasing function $f$, it holds that $f$ is submodular if and only if
$\gamma = 1$, and $f$ is supermodular if and only if $\alpha = 0$.
In addition, for this case we have that $\gamma$ and $\alpha$ lie in $[0,1]$.

With $\gamma$ and $\alpha$, the following proposition describes a guarantee
for the approximation performance of the greedy algorithm.
\begin{proposition}[\! \cite{Bian2017}]\label{prop:Bian}
Let $f$ be a nonnegative nondecreasing set function with
submodularity ratio $\gamma\in[0,1]$ and curvature $\alpha\in[0,1]$.
Then, Algorithm \ref{alg:greedy} enjoys the following approximation guarantee
for solving Problem \eqref{prob:maxf}:
\begin{align}
 f(\mathcal{S}^{\text{g}})-f(\emptyset)
 \geq \frac{1}{\alpha}\left(1-\e^{-\alpha\gamma}\right)
 \left(f(\mathcal{S}^*)-f(\emptyset)\right).
 \label{Bian_bound}
\end{align}
\end{proposition}
\medskip

Proposition~\ref{prop:Bian} generalizes the classical result,
Proposition~\ref{prop:Nemhauser}, to a case for a class of non-submodular functions.
We emphasize that for a submodular function with a small curvature, i.e.,
when $\gamma =1$ and $\alpha$ is close to 0, Proposition~\ref{prop:Bian} gives a
tighter bound than that by Proposition~\ref{prop:Nemhauser}.

Finally, we provide the concept of approximate submodularity \cite{Chamon2016}.
Although in \cite{Chamon2016} the authors use $\alpha$ to represent the degree of submodularity,
here we use $\beta$ to avoid confusion with the curvature.

\begin{definition}[\! \cite{Chamon2016}]\label{def:beta-submodularity}
A set function $f:2^{\mathcal{V}}\to \mathbb{R}$ is called $\beta$\emph{-submodular}
if $\beta\geq0$ is the largest number for which it holds that
\begin{align}
 f(\mathcal{S}_1\cup\left\{\omega\right\}) - f(\mathcal{S}_1) \geq
 \beta \left\{f(\mathcal{S}_2\cup\left\{\omega\right\}) - f(\mathcal{S}_2)\right\}
\end{align}
for all subsets $\mathcal{S}_1,\mathcal{S}_2$
($\mathcal{S}_1\subseteq\mathcal{S}_2\subseteq\mathcal{V}$) and for all
$\omega\in\mathcal{V}\backslash\mathcal{S}_2$.
\end{definition}
\medskip

The following proposition describes the relation between the submodularity ratio
and $\beta$-submodularity and is referred in the proof of the main result.

\begin{proposition}[\!\cite{Chamon2019}]\label{prop:beta-gamma}
Let $f$ be $\beta$-submodular with submodularity ratio $\gamma$.
Then, $\beta\leq\gamma$.
\end{proposition}

\section{Performance analysis of greedy solutions in the sensor placement problem}
\label{sec:main}
Let us now turn to performance guarantees for the sensor placement
problem~\eqref{prob:orig} by using Proposition~\ref{prop:Bian}.
First, to conform Problem~\eqref{prob:orig} to the maximization problem
\eqref{prob:maxf} in the previous section, let
\begin{align}
 f: 2^{\mathcal{I}}\to\mathbb{R},\quad
 f(\mathcal{S}):=-J(\mathcal{S})+J(\emptyset).
 \label{biased_MSE}
\end{align}
In the right-hand side, the second term $J(\emptyset)$ is added for normalization
as $f(\emptyset) = 0$.
Then, the sensor placement problem can be written as
\begin{subequations}\label{prob:maxfJ}
\ifshowonlyref
 \noeqref{maxfJa,maxfJb}
\fi
 \begin{alignat}{2}
 &\maximize_{\mathcal{S}\subseteq\mathcal{I}}&\quad &f(\mathcal{S}),\label{maxfJa}\\
 &\text{subject to} &\quad &|\mathcal{S}|\leq s.\label{maxfJb}
 \end{alignat}
\end{subequations}

Regarding Problem~\eqref{prob:maxfJ},
we seek to find the submodularity ratio $\gamma$ and the curvature $\alpha$ of $f$.
However, finding the exact values satisfying \eqref{gamma_def} and
\eqref{alpha_def} is computationally expensive.
Therefore, our goal is to bound them with a low computational load.

Let us introduce the following notations:
\begin{align}
 \gammamin &:= \frac{\lambdamin(L)}{\lambdamax(L+U_{\mathcal{I}})},\\
 \alphamax &:= 1-\frac{\left\{\lambdamin(L)\right\}^2}
      {\left\{\lambdamax\left(L+U_{\mathcal{I}}\right)\right\}^2}.
\end{align}
Here, $\lambdamin(\cdot)$ and $\lambdamax(\cdot)$ denote
the minimum and the maximum eigenvalues of the matrix, respectively.

We are now ready to state the main theorem.
\begin{theorem}\label{th:main}
The set function $f$ defined in \eqref{biased_MSE} is nondecreasing
and its submodularity ratio $\gamma$ and curvature $\alpha$ satisfy the
following inequalities:
\begin{align}
 \gamma \geq \gammamin > 0,\quad
 \alpha \leq \alphamax < 1.
\end{align}
\end{theorem}
\medskip

Theorem~\ref{th:main} leads us to obtain a guarantee for the greedy solution
$\mathcal{S}^{\text{g}}$ for Problem~\eqref{prob:maxfJ}.
\begin{corollary}\label{cor:approxguarantee}
 Consider Problem \eqref{prob:maxfJ}.
For the solution $\mathcal{S}^{\text{g}}$ obtained by Algorithm \ref{alg:greedy}
and the optimal solution $\mathcal{S}^*$, the following inequalities hold:
\begin{align}
 f(\mathcal{S}^{\text{g}})
 &\geq \frac{1}{\alpha}\left(1-\e^{-\alpha\gamma}\right)f(\mathcal{S}^*)\\
 &\geq  \frac{1}{\alphamax}\left(1-\e^{-\alphamax\gammamin}\right)f(\mathcal{S}^*).
 \label{approxguarantee}
\end{align}
\end{corollary}
\medskip

Note that $\gammamin$ and $\alphamax$ can be computed in polynomial time in
$|\mathcal{I}|$.
Therefore, the far right-hand side of \eqref{approxguarantee} provides a
feasible performance guarantee.
The closer the coefficient $(1-\e^{-\alphamax\gammamin})/\alphamax$ is to 1,
the smaller the guaranteed gap between the optimum and the value for
the greedy solution becomes.
In Section \ref{sec:comparison}, we illustrate how much the coefficient is
with numerical examples.

We should not ignore that Corollary~\ref{cor:approxguarantee} bounds the
deviation on $f$ rather than the original cost function $J$.
When the cost for the empty set $J(\emptyset)$ is large, the guarantee
with respect to $J$ can be conservative.
This has been pointed out in \cite{Guo2019} and an improved algorithm has been proposed.
\medskip

Before providing the proof of Theorem~\ref{th:main}, we introduce three
lemmas in linear algebra from \cite{Bernstein2018}, which will be used
in the proof.
\begin{lemma}\label{lem:invineq}
For any $A, B \succ 0$, if $A\preceq B$ then $A^{-1} \succeq B^{-1}$.
\end{lemma}
\begin{lemma}\label{lem:eigofinv}
Let $A$ be a nonsingular matrix and $\lambda$ be an eigenvalue of $A$.
Then, $1/\lambda$ is an eigenvalue of $A^{-1}$.
\end{lemma}
\begin{lemma}\label{lem:minmaxeig}
Let $A,B$ be Hermitian matrices.
It holds that
\begin{align}
 \lambdamin(A) + \lambdamin(B) &\leq \lambdamin(A+B)\\
  &\leq \lambdamin(A)+\lambdamax(B),\\
 \lambdamax(A) + \lambdamin(B) &\leq \lambdamax(A+B)\\
  &\leq \lambdamax(A)+\lambdamax(B).
\end{align}
\end{lemma}

\paragraph*{Proof of Theorem~\ref{th:main}}
The proof consists of two steps.
First, we show that $f$ is nondecreasing.
From Lemma~\ref{lem:Us}, we have
\begin{align}
 U_{\mathcal{S}_1} \preceq U_{\mathcal{S}_2}
 \label{U1leqU2}
\end{align}
for all sets $\mathcal{S}_1 \subseteq \mathcal{S}_2 \subseteq \mathcal{I}$.
Since $L\succ 0$, \eqref{U1leqU2} implies that
$L+U_{\mathcal{S}_1}\preceq L+U_{\mathcal{S}_2}$.
By this matrix inequality and Lemma~\ref{lem:invineq}, it holds that
$ \left(L+U_{\mathcal{S}_1}\right)^{-1}\succeq
  \left(L+U_{\mathcal{S}_2}\right)^{-1}$,
and thus
\begin{align}
 \tr\left[\left(L+U_{\mathcal{S}_1}\right)^{-1}\right]
 \! = \!J(\mathcal{S}_1)
 \geq \tr\left[\left(L+U_{\mathcal{S}_2}\right)^{-1}\right]
 \! = \!J(\mathcal{S}_2).
\end{align}
Accordingly, we have that $f(\mathcal{S}_1) \leq f(\mathcal{S}_2)$,
which concludes the first step.

In the second step, we evaluate the submodularity ratio $\gamma$ and
the curvature $\alpha$ of $f$.
From \cite{Chamon2017}, we have that $f$ is $\beta$-submodular and
\begin{align}
 \beta \geq \frac{ \lambdamin(L) }{ \lambdamax(L+U_{\mathcal{I}}) }.
\end{align}
Thus, from Proposition \ref{prop:beta-gamma}, we have the lower bound
$\gammamin$ on $\gamma$.

To derive $\alpha\leq\alphamax$, we bound the left- and the right-hand
sides of \eqref{alpha_def} from below and above, respectively.
The left-hand side can be evaluated as
\begin{align}
& \rho_{\left\{j\right\}}(\mathcal{S}\,\backslash\left\{j\right\}\cup\Omega)\\
 &= -\tr\left[\left( L+U_{\mathcal{S}\cup\Omega} \right)^{-1}\right]
   + \tr\left[\left( L+U_{\mathcal{S}\backslash\left\{j\right\}\cup\Omega}
              \right)^{-1}\right]\\
 &= \sum_{i=1}^{p\ell}\frac{\lambda_i[L+U_{\mathcal{S}\cup\Omega}]
    - \lambda_i[L+U_{\mathcal{S}\backslash\left\{j\right\}\cup\Omega}]}
   {\lambda_i[L+U_{\mathcal{S}\cup\Omega}]
    \lambda_i[L+U_{\mathcal{S}\backslash\left\{j\right\}\cup\Omega}]}\\
 &\geq \frac{\sum_{i=1}^{p\ell} \lambda_i[L+U_{\mathcal{S}\cup\Omega}]
    - \lambda_i[L+U_{\mathcal{S}\backslash\left\{j\right\}\cup\Omega}]}
      {\lambdamax(L+U_{\mathcal{S}\cup\Omega})
       \lambdamax(L+U_{\mathcal{S}\backslash\left\{j\right\}\cup\Omega})}\\
 &\geq \frac{ \tr[U_{\left\{j\right\}}] }
       {\left\{\lambdamax(L+U_{\mathcal{I}})\right\}^2}.
 \label{alphaLHS}
\end{align}
Here, the second equality follows by Lemma \ref{lem:eigofinv} and the
inequalities hold from Lemmas \ref{lem:Us} and \ref{lem:minmaxeig}.
On the other hand, $\rho_{\left\{j\right\}}(\mathcal{S}\backslash\left\{j\right\})$
in the right-hand side of \eqref{alpha_def} is bounded from above as
\begin{align}
 \rho_{\left\{j\right\}}(\mathcal{S}\,\backslash\left\{j\right\})
 &= -\tr\left[\left( L+U_{\mathcal{S}} \right)^{-1}\right]
   \!+\!
    \tr\left[\left( L+U_{\mathcal{S}\backslash\left\{j\right\}}
             \right)^{-1}\right]\\[-5mm]
 &= \sum_{i=1}^{p\ell}\frac{ \lambda_i[L+U_{\mathcal{S}}]
    - \lambda_i[L+U_{\mathcal{S}\backslash\left\{j\right\}}] }
   {\lambda_i[L+U_{\mathcal{S}}]
    \lambda_i[L+U_{\mathcal{S}\backslash\left\{j\right\}}]}\\
 &\leq \frac{\sum_{i=1}^{p\ell} \lambda_i[L+U_{\mathcal{S}}]
    - \lambda_i[L+U_{\mathcal{S}\backslash\left\{j\right\}}]}
      {\lambdamin(L+U_{\mathcal{S}})
       \lambdamin(L+U_{\mathcal{S}\backslash\left\{j\right\}})}\\
 &\leq \frac{ \tr[U_{\left\{j\right\}}] }
       { \left\{\lambdamin(L)\right\}^2}.
 \label{alphaRHS}
\end{align}
From \eqref{alphaLHS} and \eqref{alphaRHS}, we have
\begin{align}
 \frac{\rho_{\left\{j\right\}}(\mathcal{S}\,\backslash\left\{j\right\}\cup\Omega)}
 {\rho_{\left\{j\right\}}(\mathcal{S}\,\backslash\left\{j\right\})}
 \geq 1-\alphamax.
\end{align}

Finally, it holds that $\gammamin>0$ and $\alphamax < 1$ since $L\succ 0$.
\hfill\QED

\section{Evaluation of the derived guarantees}\label{sec:comparison}
In this section, we discuss how the approximation guarantee given in
Corollary~\ref{cor:approxguarantee} varies depending on the system model
and compare with the existing results in the literature.
We may call the ratio $f(\mathcal{S}^{\text{g}})/f(\mathcal{S}^*)$
the \emph{approximation ratio} of Algorithm~\ref{alg:greedy}
for Problem~\eqref{prob:maxfJ}.
As we have seen in \eqref{approxguarantee}, the lower bound on the approximation
ratio followed by Corollary~\ref{cor:approxguarantee} is
\begin{align}
 \frac{1}{\alphamax}\left(1-\e^{-\alphamax\gammamin}\right).\label{ARkohara}
\end{align}

\begin{figure}[tb]
 \centering
 \includegraphics[width=85mm]{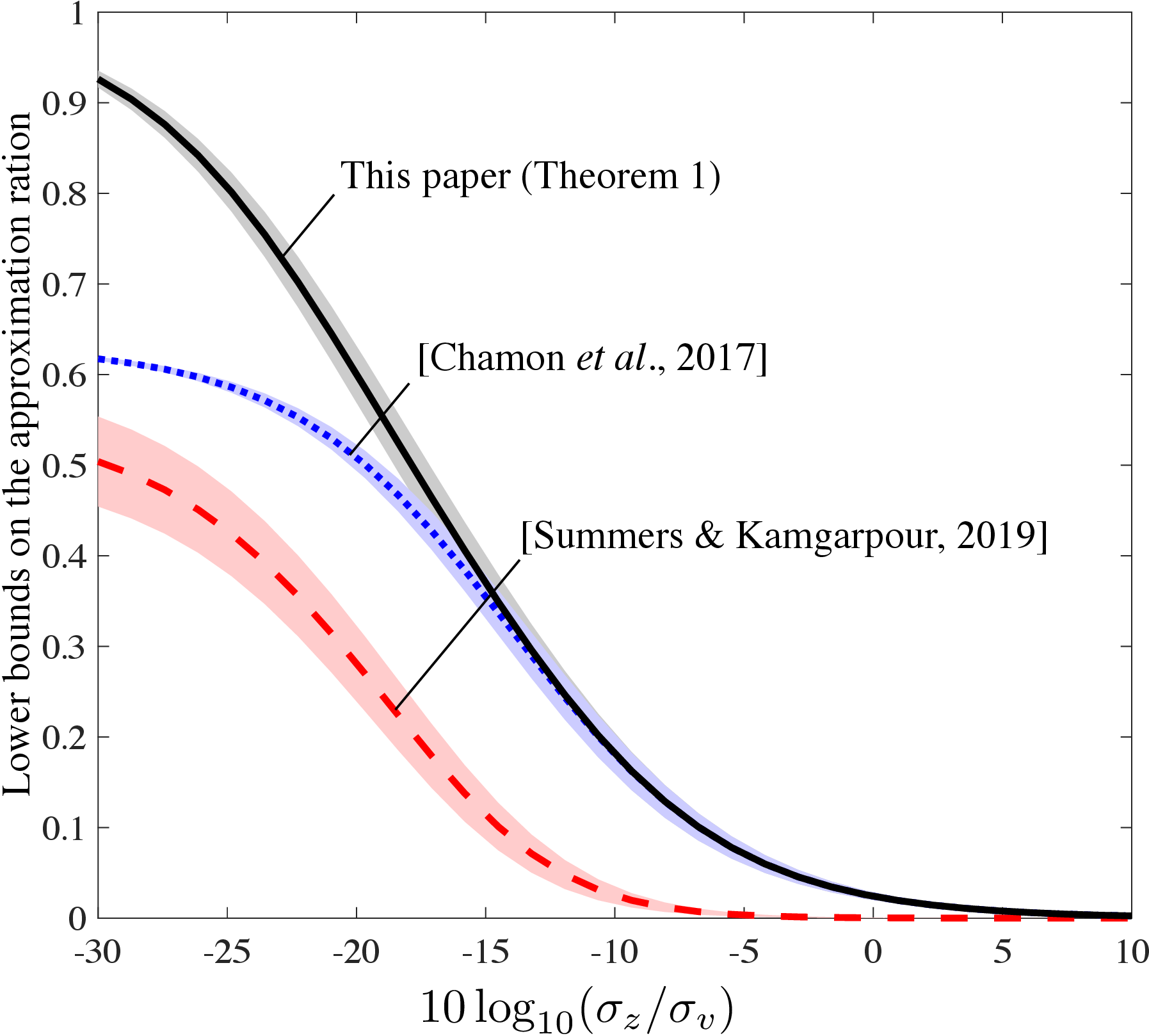}
 \caption{Lower bounds on the approximation ratio versus the noise variance:
 The bounds followed by
 Corollary~\protect\ref{cor:approxguarantee} (black solid),
 by Chamon \emph{et al.}\ \protect\cite{Chamon2017} (blue dotted),
 and by Summers and Kamgarpour \protect\cite{Summers2019a} (red dashed).
 Each line represents the mean $\pm$ the standard deviation of 1000 simulations
 with random stable $A$.
 }
 \label{fig:AR}
\end{figure}

To make the discussion simple, suppose that $C=I_n$ ($n>1$) in \eqref{system1b}
and the covariance matrix of $x_0$ and $w_k$ are represented by a single
parameter $\sigma_z$ as
$X_0 = W = \diag(\sigma_z^2,\dots,\sigma_z^2)\in\mathbb{R}^{n\times n}$.
Moreover, suppose also that
$V = \diag(\sigma_v^2,\dots,\sigma_v^2)\in\mathbb{R}^{p\times p}$.
For such a case, we have from Lemma~\ref{lem:minmaxeig} that
\begin{align}
 \gammamin &\geq
 \frac{1}
 {\left\{1 + \lambdamax(\Phi^\top \Phi)\, \sigma_v^{-2}/\sigma_z^{-2} \right\}^{2}}.
\end{align}
Similarly, it follows that
 \begin{align}
 \alphamax \leq
 1-  \frac{1}
   {\left\{ 1 + \lambdamax(\Phi^\top \Phi)\,\sigma_v^{-2}/\sigma_z^{-2}\right\}^2 }.
\end{align}
In light of these inequalities, we illustrate the bound \eqref{ARkohara} on
the approximation ratio versus $\sigma_z^2/\sigma_v^2$ numerically.
Consider the system with $n=50$ and the observation period is taken as $\ell = 10$.
The matrix $A$ in \eqref{system1a} is randomly chosen so that $A$ is Schur stable.
We fix $\sigma_v^2$ as 1 and take $\sigma_z^2$ so that $\sigma_z^2/\sigma_v^2$
varies from $-30$ to $10$ dB.
In Fig.~\ref{fig:AR}, the black solid line represents the mean of the bounds
\eqref{ARkohara} for 1000 random matrices $A$.
The shaded area around the line illustrates the standard deviation.
We see that the bound crosses $0.5$ at around $\sigma_z^2/\sigma_v^2=-20$ dB,
that is, our result guarantees that the value obtained by the greedy solution is
more than half of the optimum.
For a smaller $\sigma_z^2/\sigma_v^2$ around $-30$ dB, the bound ensures that
$f(\mathcal{S}^{\text{g}})$ becomes more than $92$\%
On the other hand, when the variance of the process disturbance is relatively
large, the derived bound becomes small and does not make sense.
We have omitted for the cases $\sigma_z^2/\sigma_v^2 > 10$ dB since such a case
is less interesting.

\addtolength{\textheight}{-5cm}

Here, we make comparisons with the derived bound \eqref{ARkohara} and those
in the existing work.
In \cite{Chamon2017}, the authors study the sensor placement and the filtering
problems minimizing the means square error.
They have evaluated the objective function by using $\beta$-submodularity
in Definition \ref{def:beta-submodularity}.

The result in \cite{Chamon2017} on the sensor placement problem can be summarized
as follows.
\begin{proposition}[\!\cite{Chamon2017}]\label{prop:chamon}
 Consider Problem~\eqref{prob:maxfJ}.
For the solution $\mathcal{S}^{\text{g}}$ by Algorithm~\ref{alg:greedy}
and the optimal solution $\mathcal{S}^*$, it holds that
\begin{align}
 &f(\mathcal{S}^{\text{g}}) \geq (1-\e^{-\beta}) f(\mathcal{S}^*),\quad
 \beta \geq \frac{\lambdamin(L)}{\lambdamax\left(L+U_{\mathcal{I}}\right)}.
\end{align}
\end{proposition}
\medskip

Notice that the right-hand side of the first inequality increases with respect to $\beta$.
If $\beta = 1$, the inequality coincides with the bound given in
Proposition~\ref{prop:Nemhauser}, which is valid for submodular functions.
Furthermore, since the lower bound on $\beta$ is equal to $\gammamin$,
the feasible bound
\begin{align}
 1-\e^{- \lambdamin(L) / \lambdamax\left(L+U_{\mathcal{I}}\right) } \label{ARchamon}
\end{align}
followed by Proposition~\ref{prop:chamon} can be considered as a special of
\eqref{ARkohara} where $\alpha=1$.
When the curvature $\alpha<1$, \eqref{ARkohara} is greater than \eqref{ARchamon}
and thus gives a less conservative guarantee.

To confirm the bound \eqref{ARchamon}, we have performed the same simulation as for
\eqref{ARkohara}.
In Fig.~\ref{fig:AR}, we plot the bound \eqref{ARchamon} as a blue dotted line.
We see that for a smaller value of $\sigma_z^2/\sigma_v^2$, \eqref{ARkohara} is
considerably tighter than \eqref{ARchamon}.
It should be emphasized that the blue dotted line does not exceed
$1-\e^{-1}\approx 0.63$ even though it increases as $\sigma_z^2/\sigma_v^2$
becomes small.
On the contrary, the black solid line reaches $0.92$ at $-30$ dB.
This point illustrates the advantage of our approach.

We next introduce the result in \cite{Summers2019a}.
The authors consider a linear system with no process disturbance $w_k$ and
an unknown but deterministic initial state $x_0$.
For such a system, the actuator placement to minimize the
average energy required to move the state is studied.
The objective function is described by the infinite-horizon controllability Gramian
and is analyzed based on the submodularity ratio and curvature.
By applying the result in \cite{Summers2019a} to the sensor placement problem
\eqref{prob:maxfJ}, we have the following proposition.
\begin{proposition}\label{prop:summers}
  Consider Problem~\eqref{prob:maxfJ}.
 Then, $f$ is nondecreasing and the submodularity ratio $\gamma$ and curvature
$\alpha$ of $f$ are bounded as
\begin{align}
 &\gamma \geq \gammamin':=
  \frac{\min_{\omega\in\mathcal{I}}\tr(U_{\left\{\omega\right\}\!})
       \left\{\min_{\omega\in\mathcal{I}}\lambdamin(L\!+\!U_{\left\{\omega\right\}\!})
       \right\}^2}
      {\max_{\omega\in\mathcal{I}} \tr(U_{\left\{\omega\right\}})
       \left\{ \lambdamax(L+U_{\mathcal{I}}) \right\}^2 },\\
 &\alpha \leq \alphamax' := 1 - \gammamin'.
\end{align}
\end{proposition}
\medskip

By following the same discussion for the derivation of Corollary~\ref{cor:approxguarantee},
we have
\begin{align}
 \frac{1}{\alphamax'}(1-\e^{-\alphamax'\gammamin'})\label{ARsummers}
\end{align}
as a lower bound on the approximation ratio.
Again, we plot \eqref{ARsummers} with respect to $\sigma_z^2/\sigma_v^2$
as the red dashed line in Fig.~\ref{fig:AR}.
We find that the bound \eqref{ARkohara} is greater than \eqref{ARsummers}
for all noise variances.

\section{Conclusion}\label{sec:conclusion}
In this paper, we have considered to minimize the state estimation MSE in the
smoothing problem with respect to the sensor placement.
For evaluating the objective value corresponding to the greedy solution,
we have analyzed the submodularity ratio and the curvature of the MSE function
and have derived bounds on these quantities.
By using the obtained bounds, a performance guarantee for the greedy algorithm
has been given.
Through numerical simulations, we have shown that our performance guarantee exceeds
the classical bound $1-\e^{-1}$
and becomes tighter than those in the existing work for a class of systems.

\emph{Acknowledgment}: The authors would like to thank the anonymous reviewers for their
helpful comments, especially on the proof of Theorem~\ref{th:main}.

\end{document}